\documentstyle[12pt,epsfig]{article}
\def\be{\begin{equation}} \def\ee{\end{equation}} \def\bea{\begin{eqnarray}}
\def\eea{\end{eqnarray}} \def\nnb{\nonumber}

\begin{document}

\hfill{TRI-PP-05-25, \ USC(NT)-05-06}


\begin{center}
\vskip 10mm
{\Large\bf Neutron-Neutron Fusion}
\vskip 10mm
{\large Shung-ichi Ando $^{a,}$\footnote{E-mail:sando@triumf.ca} 
and 
Kuniharu Kubodera$^{b,}$\footnote{E-mail:kubodera@sc.edu}}
\\
\vskip 10mm
{\it $^a$ Theory Group, TRIUMF, 4004 Wesbrook Mall, Vancouver,
B.C. V6T 2A3, Canada}\\
\vskip 3mm
{\it $^b$ Department of Physics and Astronomy, 
University of South Carolina, Columbia, SC 29208, USA}
\end{center}

\vskip 7mm

The neutron-neutron fusion process, 
$nn\to de\nu$, at very low neutron energies
is studied in the framework of 
pionless effective field theory 
that incorporates dibaryon fields.
The cross section and electron energy spectrum
for this process are calculated up 
to next-to-leading order.
We include the radiative corrections of ${\cal{O}}(\alpha)$
calculated for the one-body transition amplitude.
The precision of our theoretical estimates
is found to be governed essentially by the accuracy
with which the empirical values
of the neutron-neutron scattering length and effective range
are currently known.
Also discussed is the precision of 
theoretical estimates of the transition rates
of related electroweak processes in few-nucleon systems.

\newpage

\noindent
{\bf 1. Introduction}

The ultra-high-intensity neutron-beam facilities
currently under construction at, e.g., 
the Oak Ridge National Laboratory and the J-PARC 
are expected to bring great progress 
in high-precision experiments
concerning the fundamental properties of the neutron 
and neutron $\beta$-decay.
The planned experiments of great importance
include the accurate determination of 
the neutron electric dipole moment,
and high-precision measurements of the
lifetime and the correlation coefficients 
for neutron $\beta$-decay.
Besides these experiments that focus on 
the properties of a single neutron, 
one might be tempted to consider
processes that involve
the interaction of two free neutrons.
An example is the direct observation
of free neutron-neutron scattering,
which would allow the model-independent 
determination of the neutron-neutron 
scattering length $a_0^{nn}$ and 
effective range $r_0^{nn}$.
These quantities play an important role
in high-precision calculations of 
low-energy electroweak transitions
involving the $nn$ channel (see below).
Their accurate values are also important
in the study of isospin symmetry breaking effects
in the strong interactions~\cite{sat-pr89,mns-pr90}.
Up to now, however, 
because of the lack of {\it free} neutron targets,
information on $a_0^{nn}$ and $r_0^{nn}$ 
has been obtained 
by analyzing processes such as 
$\pi^-d\to nn\gamma$ or $nd\to nnp$;    
for the current status of these experiments, 
see, e.g., 
Refs.~\cite{hetal-plb98,getal-prl99,hetal-prl00}.
Although there has been considerable improvement
in these analysis, they still contain
rather significant theoretical uncertainties.
For instance, the treatments of the exchange currents
and higher partial waves 
in the $\pi^-d\to nn\gamma$ process~\cite{gp-05}
are open to further examinations;
similarly, the treatment of the three-body interactions 
in the $nd\to nnp$ reaction is yet to be settled.
Therefore, the observation of free neutron-neutron 
scattering is an interesting alternative,
even though its extremely low counting rate
is certainly a major obstacle.\footnote{
There exists a project
to detect the scattering of two free neutrons 
using reactor-generated neutrons~\cite{cetal-jpg04}.}

As another process involving two free neutrons, 
we consider here the neutron-neutron fusion reaction
\bea
n+n\to d+ e^-+\bar{\nu}_e\, ,
\eea
for neutrons of very low energies
such as the ultra-cold neutrons and
thermal neutrons.
The experimental observation of this reaction 
does not seem to belong to the near future   
but, in view of the existing strong thrust
for pursuing experiments that take advantage
of ultra-high-intensity neutron beams, 
it may not be totally purposeless
to make a theoretical study of $nn$-fusion,
and this is what we wish to do here.
It is worth noting that, for very low energy neutrons,
the maximum energy $E_e^{max}$ of the outgoing electron
is $E_e^{max} \simeq B+ \delta_N \simeq 3.52$ MeV,
where $B$ is the deuteron binding energy  
and $\delta_N $ is the mass difference 
between the neutron and proton,
$\delta_N=m_n -m_p$.
This value of $E_e^{max}$ is significantly 
larger than the maximum energy of electrons 
from neutron $\beta$-decay, 
$E_{e,\beta\mbox{-}decay}^{max}
\simeq \delta_N \simeq 1.29$ MeV.
Thus the $nn$-fusion electrons with energies
larger than $\delta_N$ 
are in principle distinguishable 
from the main background 
of the neutron $\beta$-decay electrons.

Our study is based on 
low-energy effective field theory (EFT). 
For a system of mesons and baryons
(without or with external probes),
EFT provides a systematic perturbative expansion scheme 
in terms of $Q/\Lambda$,
where $Q$ denotes a typical momentum scale 
of the process under consideration,
and $\Lambda$ represents a cutoff scale 
that characterizes 
the EFT lagrangian~\cite{w-97}.
In a higher order effective lagrangian,
there appear unknown parameters, 
so-called low energy constants (LECs),
which represent the effects of high-energy physics
that has been integrated out.
In many cases, these LECs cannot be determined
by the symmetries of the theory alone and hence 
need to be fixed by experiments.
Once all the relevant LECs are fixed,
the EFT lagrangian represents a complete
(and hence model-independent) lagrangian 
to a given order of expansion.

Many studies based on EFT
have been made of nuclear processes 
in few-nucleon systems;
for reviews see, e.g.,
Refs.~\cite{betal-00,bk-arnps02,kk-04} 
and references therein.
These studies have used various versions 
of EFT that differ primarily in the treatment
of the following points.
(1) Choice between the 
Weinberg counting scheme~\cite{w-plb90} and
the KSW counting scheme~\cite{ksw-plb98};
(2) Inclusion or exclusion of 
the pion field 
as an active degree of freedom~\cite{crs-npa99};
(3) The presence of
an abnormally low energy scale
reflecting the large scattering lengths
in the nucleon-nucleon interaction
is handled by introducing
an expansion of the inverse 
of the scattering amplitude~\cite{ksw-plb98},
or by introducing 
the effective dibaryon fields
(see, e.g., Ref.~\cite{bs-npa01}).
(For convenience, we refer to the former
as the $1/{\cal{A}}$ expansion method,
and the latter as dEFT.)
For the reasons to be explained below,
we use here pionless dEFT.

The $nn$-fusion process
of our concern here is closely related 
to the $pp$-fusion reaction [$pp\to de^+\nu$],
and the $\nu d$ reactions [$\nu d\to ppe(np\nu)$,
 ${\bar{\nu}} d\to nne^+(np{\bar{\nu}})$], 
which have been studied extensively
because of their importance in
astrophysics and neutrino physics;
see, e.g., 
Refs.~\cite{petal-aj98,kr-npa99,
bc-plb01,petal-prc03}
for $pp$-fusion,
and 
Refs.~\cite{nsgk-prc01,netal-npa02,
bc-npa00,bck-prc01,aetal-plb03}
for the $\nu d$ reactions.
In an EFT treatment of these reactions,
there occurs a LEC
that controls the strength
of contact two-nucleon-axial current coupling.
This LEC is referred to as $\hat{d}^R$ 
in Weinberg-scheme calculations~\cite{petal-prc03}
and as $L_{1A}$ 
in KSW-scheme calculations~\cite{bck-prc01}.
Park {\it et al.} have developed 
a useful hybrid approach called EFT* to determine 
$\hat{d}^R$~\cite{petal-prc03}.\footnote{
This approach is also called MEEFT 
(more effective EFT); for a review, 
see, e.g., Ref.~\cite{kk-04}.}
In EFT*, the nuclear transition amplitude
for an electroweak process is calculated
with the use of the relevant transition operator
derived from EFT along with the nuclear
wave functions obtained
from a high-precision phenomenological 
nucleon-nucleon potential.\footnote{
For discussion of some formal aspects of EFT*,
see Ref.~\cite{n-04}.}
An advantage of EFT* is that it can be
used for light complex nuclei (A=3,4, \ldots)
with essentially the same accuracy 
and ease as for the two-nucleon systems.
The use of EFT* enabled Park {\it et al.} 
to determine $\hat{d}^R$ with good accuracy
from the tritium 
$\beta$-decay rate~\cite{petal-prc03}.\footnote{
Attempts to determine $L_{1A}$
can be found in, e.g.,
Refs.~\cite{bcv02,chhr-prc03,ds-npa04};
for the possible use of 
$\mu d$ capture
to determine $\hat{d}^R$ or $L_{1A}$,
see Refs.~\cite{aetal-plb02,cijl-05,k-03}.  }
This result allowed parameter-free calculations
of the $pp$-fusion and $Hep$-reaction rates
\cite{petal-prc03}, and 
the $\nu d$-reaction cross 
sections~\cite{aetal-plb03}.
It is certainly possible to extend
this EFT* calculation to the $nn$-fusion process,
but we do not attempt it here.
Instead, we employ pionless dEFT
with motivations to be described below. 

As is well known,
the EFT treatment of low-energy nucleon-nucleon 
scattering is complicated by the existence
of a large length scale associated with
the very-weakly bound state (deuteron)
in the $^3S_1$-$^3D_1$ channel, or
with the near-threshold resonance (virtual state,
or singlet deuteron) in the $^1S_0$ channel. 
The $1/{\cal{A}}$-expansion method was
proposed to cope with this difficulty.
This method, however, exhibits
rather slow convergence in the deuteron channel,
for which a typical expansion parameter is
$(\gamma\rho_d) \simeq 0.4\sim 1/3$,
where 
$\gamma\simeq 45.7$ MeV and 
$\rho_d\simeq 1.764$ fm.
Thus, to achieve 1 \% or 3 \% accuracy, 
one must go up 
to next-to-next-to-next-to-next-to-leading order (N$^4$LO) 
or next-to-next-to-next-to-leading(N$^3$LO) order
($(1/3)^4 \sim 1$ \% or $(1/3)^3\sim 3$\%).
These N$^4$LO and N$^3$LO calculations are 
formidably challenging even 
in the pionless cases.\footnote{   
It has been suggested 
by Rho and other authors 
that much better convergence is achieved
by adjusting the deuteron wave function 
to fit the asymptotic 
$S$-state normalization constant, 
$Z_d=\gamma\rho_d/
(1-\gamma\rho_d)$~\cite{r-99,pc-npa00,prs-plb00}.}
Another method to resolve the difficulty associated
with the existence of the large length scale
in nucleon-nucleon scattering is
to introduce explicit $S$-channel states,
called the ``dibaryons",
which represent the near-threshold resonance state 
for the  $^1S_0$ channel,
and the deuteron state 
in the $^3S_1$-$^3D_1$ channel~\cite{k-npb97,bk-plb98}.
(As mentioned earlier, 
we refer to an EFT that includes the dibaryon fields
as dEFT.)
Beane and Savage~\cite{bs-npa01} discussed
counting rules in dEFT
and moreover demonstrated 
the usefulness of dEFT in 
describing the electromagnetic processes
in the two-nucleon systems.
Furthermore, a recent study~\cite{ah-04} 
of the electroweak observables 
of the deuteron
shows that dEFT allows one
to achieve in a very 
transparent and economical way
the level of accuracy
that, 
in the $1/{\cal{A}}$-expansion method,
would require the inclusion of 
terms of very high orders.

The above consideration motivates us
to use dEFT in studying the $nn$-fusion process.
We limit ourselves here to the case
of very low incident neutron energy,
which makes it safe to eliminate the pion
fields and concentrate on pionless dEFT.
We calculate the total cross section 
and electron energy spectrum 
for $nn$-fusion up to NLO in dEFT.
Although the enormous difficulty of 
observing $nn$-fusion may render it unwarranted
to pursue high precision in our calculation,
we still wish to aim at a few percent accuracy
for the following reason.
As mentioned,
$nn$-fusion is closely related
to $pp$-fusion and the $\nu d$ reaction,
and the latter two processes
do require high-precision calculations.
Since our present formalism can be applied
(with essentially no change) 
to high-precision estimation 
of the $pp$-fusion and $\nu d$ cross sections,
we may use the $nn$-fusion case
to explain what is involved in
those high-precision calculations. 
It turns out (see below)
that, at the level of a few percent accuracy,
we need to treat properly: 
(1) the LEC, $l_{1A}$,
which represents the strength of 
a dibaryon-dibaryon-axial-vector ($ddA$)
interaction and which is 
associated with $\hat{d}^R$ and $L_{1A}$
discussed earlier;
(2) the radiative corrections;
(3) the influence of uncertainties 
in the currently available experimental information  
on $a_0^{nn}$ and $r_0^{nn}$.
We will show that 
main uncertainties 
in our calculation of
the low-energy $nn$-fusion cross section
come from the last item (3).

\vskip 3mm \noindent
{\bf 2. Effective lagrangian}

For low-energy processes, 
the weak-interaction Hamiltonian can be taken 
to be 
\bea
{\cal H} &=& \frac{G_FV_{ud}}{\sqrt{2}}l_\mu J^\mu\,  ,
\eea
where $G_F$ is the Fermi constant and $V_{ud}$ is the 
CKM matrix element.
$l_\mu$ is the lepton current 
$l_\mu = \bar{u}_e \gamma_\mu (1-\gamma_5)v_\nu$,
while $J_\mu$ is the hadronic current,
which we calculate here up to two-body terms
based on the effective lagrangian
of dEFT. 

We adopt the standard counting rules
of dEFT~\cite{bs-npa01}.
Introducing an expansion scale $Q<\Lambda(\simeq m_\pi)$,
we count the magnitude of spatial part of the external and loop
momenta, $|\vec{p}|$ and $|\vec{l}|$, as $Q$,
and their time components, $p^0$ and $l^0$, as $Q^2$.
The nucleon and dibaryon propagators are of $Q^{-2}$,
and a loop integral carries $Q^5$. 
The scattering lengths and effective ranges are counted as 
$Q\sim \{\gamma, 1/a_0,1/\rho_d,1/r_0\}$.
The orders of vertices and transition amplitudes 
are easily obtained by counting the numbers 
of these factors in the lagrangian
and diagrams, respectively.
As discussed below, some vertices acquire factors
like $r_0$ and $\rho_d$ after renormalization and thus 
their orders can differ from what
the above naive dimensional analysis suggests.

A pionless dEFT lagrangian may 
be written as~\cite{bs-npa01,ah-04}
\bea
{\cal L} &=& {\cal L}_N + {\cal L}_s + {\cal L}_t + {\cal L}_{st}\, ,
\eea
where ${\cal L}_N$ is a one-nucleon lagrangian, 
${\cal L}_s$ is the spin-singlet dibaryon 
lagrangian including coupling to the nucleon,
${\cal L}_t$ is the spin-triplet dibaryon 
lagrangian including coupling to the nucleon;
${\cal L}_{st}$ describes
the weak-interaction transition 
(due to the axial current)
from the $^1S_0$ dibaryon to 
the $^3S_1$ dibaryon. 
A pionless one-nucleon lagrangian 
in the heavy-baryon formalism reads
\bea
{\cal L}_N &=& 
N^\dagger \left\{ iv\cdot D -2ig_AS\cdot \Delta
+\frac{1}{2m_N}\left[(v\cdot D)^2 -D^2\right] +\cdots
\right\} N\, ,
\eea
where the ellipsis represents terms that do not appear 
in this calculation.
$v^\mu$ is the velocity vector satisfying $v^2=1$;
we choose $v^\mu=(1,\vec{0})$.
$S^\mu$ is the spin operator 
$2S^\mu = (0,\vec{\sigma})$,
while
$D_\mu = \partial_\mu 
-\frac{i}{2}\vec{\tau}\cdot \vec{\cal V}_\mu$
where $\vec{\cal V}_\mu$ 
is the external isovector vector current; 
$\Delta_\mu = 
-\frac{i}{2}\vec{\tau}\cdot \vec{\cal A}_\mu$,
where $\vec{\cal A}_\mu$ is 
the external isovector axial current.
$g_A$ is the axial-vector coupling constant,
and $m_N$ is the nucleon mass.
The terms that involve the dibaryon fields 
are given by
\bea
{\cal L}_s &=& \sigma_s s_a^\dagger \left[
iv\cdot D + \frac{1}{4m_N}[(v\cdot D)^2 -D^2] + \Delta_s \right] s_a
-y_s \left[ s_a^\dagger (N^TP_a^{(^1S_0)}N) +\mbox{\rm h.c.}
\right]\, ,
\\
{\cal L}_t &=& \sigma_t t_i^\dagger \left[
iv\cdot D + \frac{1}{4m_N}[(v\cdot D)^2 -D^2] + \Delta_t \right] t_i
-y_t \left[ t_i^\dagger (N^TP_i^{(^3S_1)}N) +\mbox{\rm h.c.}
\right]\, ,
\\
{\cal L}_{st} &=& 
- \left[ \left(\frac{r_0+\rho_d}{2\sqrt{r_0\rho_d}}\right)\, g_A
+ \frac{l_{1A}}{m_N\sqrt{r_0\rho_d}}\right] 
\left[ s_a^\dagger  t_i {\cal A}_i^a + \mbox{\rm h.c.}\right]\, ,
\label{eq;Lst}
\eea
where $s_a$ and $t_i$ are the dibaryon fields for the $^1S_0$ and
$^3S_1$ channel, respectively.
The covariant derivative for the dibaryon field is given by
$D_\mu = \partial_\mu -iC{\cal V}^{ext}_\mu$ where
${\cal V}_\mu^{ext}$ is the external vector field.
$C$ is the charge operator for the dibaryon field;
$C=0,1,2$ for the $nn$, $np$, $pp$ channel, respectively.
$\sigma_{s,t}$ is the sign factor and 
$\Delta_{s,t}$ is the mass difference between the dibaryon 
and two nucleons, $m_{s,t} = 2m_N+\Delta_{s,t}$.
$r_0$ and $\rho_d$ are the effective ranges 
for the deuteron and $^1S_0$ state, respectively.
$P_i^{(S)}$ is the projection operator 
for the $S$ = $^1S_0$ or $^3S_1$ channel;
\bea
P_a^{(^1S_0)} = \frac{1}{\sqrt8}\sigma_2\tau_2\tau_a\, , 
\ \ \ 
P_i^{(^3S_1)} = \frac{1}{\sqrt8}\sigma_2\sigma_i\tau_2\, ,
\ \ \ 
{\rm Tr}\left(P_i^{(S)\dagger}P_j^{(S)}\right)
= \frac12 \delta_{ij}\, ,
\eea
where $\sigma_i$ ($\tau_a$) is the spin (isospin) operator.

The LECs, $y_s$ and $y_t$, represent the dibaryon-$NN$ ($dNN$)
couplings in the spin-singlet and spin-triplet states,
respectively.
These LECs along with
$\Delta_{s,t}$ and $\sigma_{s,t}$
are to be determined from the effective ranges in 
the $^1S_0$ and $^3S_1$ channels.
\begin{figure}
\begin{center}
\epsfig{file=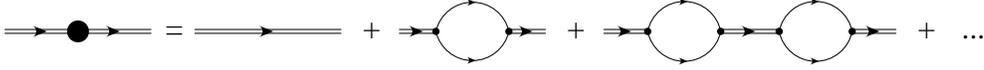,width=13cm}
\caption{
Dressed dibaryon propagator 
(double line with a filled circle)
at leading order.
A single line stands for the nucleon,
while a double line represents the bare dibaryon.}
\label{fig;NN-d-prop}
\end{center}
\end{figure}
LO diagrams for the 
dressed dibaryon propagators
are depicted in Fig.~\ref{fig;NN-d-prop}.
Since an insertion of the two-nucleon one-loop diagram does not
alter the order of the diagram, the two-nucleon bubbles 
should be summed up to infinite order.
Thus the inverse 
of the dressed dibaryon propagator in 
the center-of-mass (CM) frame reads
\bea
iD_{s,t}^{-1}(p) &=& i\sigma_{s,t} (E+\Delta_{s,t}) 
+ i y_{s,t}^2 \frac{m_N}{4\pi} (ip)
\nnb \\ &=& 
i\frac{m_Ny_{s,t}^2}{4\pi}
\left[
  \frac{4\pi\sigma_{s,t} \Delta_{s,t}}{m_Ny^2_{s,t}}
+ \frac{4\pi\sigma_{s,t}E}{m_Ny_{s,t}^2}+ip
\right]\, ,
\eea
where we have used dimensional regularization for
the loop integral and $E$ is the total energy of the
two nucleons,
$E\simeq p^2/m_N$.
\begin{figure}
\begin{center}
\epsfig{file=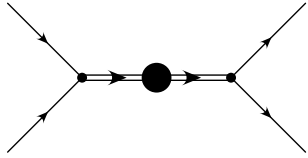,width=4cm}
\caption{
Diagram for the $S$-wave $NN$ scattering amplitude at leading order. 
The double line with a filled circle represents the 
dressed dibaryon 
propagator obtained in Fig.~\ref{fig;NN-d-prop}.
}
\label{fig;NN}
\end{center}
\end{figure}
The dressed dibaryon propagators are renormalized 
via the $S$-wave $NN$ scattering amplitudes.
The amplitudes obtained from the diagram 
in Fig.~\ref{fig;NN} 
should satisfy 
\bea
iA_{s,t} &=& (-iy_{s,t})\left(
iD_{s,t}(p)\right) (-iy_{s,t})
= \frac{4\pi}{m_N}
  \frac{i}{-\frac{4\pi \sigma_{s,t}\Delta_{s,t}}{m_Ny_{s,t}^2}
- \frac{4\pi \sigma_{s,t}}{m_Ny_{s,t}^2} p^2
- ip } \, ,
\eea
where $A_{s,t}$ is related to the $S$-wave NN scattering 
$S$-matrix via
\bea
S-1 = e^{2i\delta_{s,t}}-1 = \frac{2ip}{p\, {\rm cot}\delta_{s,t}-ip}
= i\left(\frac{pm_N}{2\pi}\right) A_{s,t}\, , 
\eea 
Here $\delta_s$ ($\delta_t$) is the phase shift 
for the $^1S_0$ ($^3S_1$) channel.
Meanwhile, effective range expansion reads
\bea
p\, {\rm cot}\delta_s = - \frac{1}{a_0} + \frac12 r_0p^2 + \cdots, 
\ \ \ 
p\, {\rm cot}\delta_t = - \gamma + \frac12 \rho_d(\gamma^2+p^2) + \cdots, 
\eea
where $a_0$ and $r_0$ are the scattering length and effective 
range for the $^1S_0$ channel and $\gamma$ is the deuteron momentum
$\gamma=\sqrt{m_NB}$ ($B$ is the deuteron binding energy) 
and $\rho_d$ is the effective range for the $^3S_1$ channel. 
Now, the above renormalization condition
allows us to relate the LECs 
to the effective-range expansion parameters. 
For the $^1S_0$-channel,
this procedure leads to $\sigma_s=-1$, 
\bea
y_s = \frac{2}{m_N}\sqrt{\frac{2\pi}{r_0}}\, ,
\ \ \ 
D_s(p) = \frac{m_Nr_0}{2}\frac{1}
{\frac{1}{a_0}+ip-\frac12 r_0p^2}\, .
\eea
For the deuteron channel, one has
$\sigma_t=-1$ and 
\bea
y_t = \frac{2}{m_N}\sqrt{\frac{2\pi}{\rho_d}}\, ,
\ \ \ 
D_t(p) = \frac{m_N\rho_d}{2}
\frac{1}{\gamma+ip-\frac12 \rho_d(\gamma^2+p^2)}
= \frac{Z_d}{E+B} +\cdots \, ,
\label{eq;Dt}
\eea
where $Z_d$ is the wave function normalization factor
of the deuteron at the pole $E=-B$,
and the ellipsis in Eq.~(\ref{eq;Dt})
denotes corrections that are finite or vanish at 
$E=-B$. 
Thus one has \cite{bs-npa01}
\bea
Z_d = \frac{\gamma\rho_d}{1-\gamma\rho_d}\, .
\eea
This $Z_d$ is equal to 
the asymptotic $S$-state normalization constant
discussed in Introduction.
It is to be noted  
that the order of the
LECs $y_{s,t}$ is now of $Q^{1/2}$, and 
the deuteron state is described 
by the dressed dibaryon propagator
that contains two-nucleon loops as well as 
the bare $^3S_1$ dibaryon.

The $ddA$ vertex in Eq.~(\ref{eq;Lst}) contains 
a LEC $l_{1A}$, which is associated with 
the LEC, $\hat{d}^R$ or $L_{1A}$,
appearing in the contact-type two-nucleon-axial-vector
vertex. 
It is not obvious how to relate $l_{1A}$ to
$\hat{d}^R$ or $L_{1A}$, because 
the dimensions of these LECs 
are different; $l_{1A}$ is a dimensionless quantity,
whereas, for instance, $L_{1A}$ is measured in units of fm$^{3}$ 
because of two more baryon fields involved in the vertex.
However, a relation between $l_{1A}$ and $L_{1A}$ 
is discussed in Refs.~\cite{ds-npa04,cijl-05}.
We employ here the assumption 
proposed in Ref.~\cite{ah-04} 
that $l_{1A}$ involves 
both LO and subleading-order parts.
The LO part is fixed from the  
one-body $NNA$ interaction vertex,
which is proportional to $g_A$
and the factor 
$\left(\frac{r_0+\rho_d}{2\sqrt{r_0\rho_d}}\right)\simeq 1.024$,
which has been introduced so as to reproduce the result 
of effective range theory.
The subleading term $l_{1A}$ represents a two-body 
interaction and its value is fixed by using the ratio
of the two- and one-body amplitudes (see below).

\vskip 3mm \noindent
{\bf 3. Cross section and numerical results}

\begin{figure}
\begin{center}
\epsfig{file=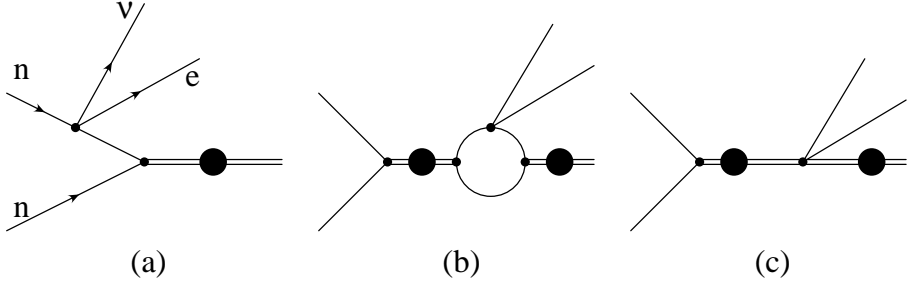,width=12cm}
\caption{Diagrams for neutron-neutron fusion,
$nn\to de\nu$, up to next-to leading order.}
\label{fig;nn}
\end{center}
\end{figure}
We calculate the $nn$-fusion amplitude 
by adding the contributions
from diagrams (a), (b) and (c) in Fig.~\ref{fig;nn}.
Since the initial two neutrons are in the $^1S_0$ state,
the dressed dibaryon propagator 
in diagrams (b) and (c) is limited to be $D_s(p)$
(the contribution from $D_t(p)$ is highly suppressed 
due to the orthogonality between the initial scattering and
final bound $^3S_1$ states).
Meanwhile, the final deuteron state 
is described by the wavefunction normalization factor 
$\sqrt{Z_d}$.
Thus we have the $nn$-fusion amplitude
\bea
A^{(a+b+c)}(^1S_0) &=& \vec{\epsilon}^*_{(d)}\cdot \vec{\epsilon}_{(l)}
G_FV_{ud}
\sqrt{\frac{2\pi \gamma}{1-\gamma\rho_d}}
\frac{2}{m_N}
\frac{a_0^{nn}g_A}{E_e^{max}}
\nnb \\ && \times \left[
\sqrt{m_NE_e^{max}} -\frac{1}{a_0^{nn}}
-\frac14(r_0^{nn}+\rho_d) m_NE_e^{max}
-\frac{E_e^{max}}{2g_A}l_{1A}
\right] \, ,
\label{eq;Amp}
\eea
where $\vec{\epsilon}_{(d)}^*$ is the spin polarization vector
of the deuteron and $\vec{\epsilon}_{(l)}$ is the spatial 
part of the lepton current $l^\mu =(\epsilon^0_{(l)},\vec{\epsilon}_{(l)})$.
Note that the amplitude in Eq.~(\ref{eq;Amp})
is proportional to $a_0^{nn}$ and $g_A$.
It also depends on $l_{1A}$.
We remark that 
the amplitude obtained above is similar to that for
low-energy $np\to d\gamma$ capture 
calculated in the effective-range expansion approach
and in an NLO dEFT calculation;
the $np\to d\gamma$ process involves
the same partial waves, the initial $^1S_0$ wave 
and the final $^3S_1$ deuteron state.
Thus, by changing  
$\sqrt{m_N E_e^{max}}$ in the bracket
in Eq.~(\ref{eq;Amp}) to $\sqrt{m_NB}$,
one obtains an expression analogous
to the amplitude for $np\to d\gamma$
(see, e.g., Eq.~(39) in Ref.~\cite{ah-04}). 

The differential cross section for $nn$-fusion
is now easily obtained.
We include here the Fermi function,
which describes the Coulomb interaction 
between the out-going electron and the deuteron.
Furthermore, we take into account
the radiative corrections of ${\cal{O}}(\alpha)$
calculated for the one-body 
transition diagrams~\cite{aetal-plb04};
here $\alpha$ is the fine structure constant.
These effects need to be incorporated
in order to achieve accuracy better than 1 \% 
in the calculated cross section.
We then arrive at\footnote{
We have used the relation
$\sum_{spin} |\vec{\epsilon}_{(d)}^*\cdot \vec{\epsilon}_{(l)}|^2 = 
6-2\beta y$,
where $\beta=|\vec{p}_e|/E_e$ and $y=\hat{p}_e\cdot\hat{p}_\nu$.
}
\bea
\frac{d\sigma}{dE_e} &=&
6 p_eE_e(E_e^{max}-E_e)^2
F(Z,E_e) \left[ 1+
\frac{\alpha}{2\pi} \delta^{(1)}_\alpha
\right]
\nnb \\ && \times
\frac{1}{v}
\frac{G_V'^2}{\pi^2}
\left(\frac{a_0^{nn}g_A}{\gamma_{nn}}\right)^2
\frac{\gamma}{1-\gamma\rho_d}
\left[
1
-\frac{1}{\gamma_{nn}a_0^{nn}}
-\frac14 (r_0^{nn}+\rho_d) \gamma_{nn}
-\frac{\gamma_{nn}l_{1A}}{2m_Ng_A} 
\right]^2 \, ,
\label{eq;cross-section}
\eea
where $F(Z,E_e)$
is the Fermi function defined by 
$F(Z,E_e)=x/(1-exp(-x))$ with $x=2\pi \alpha Z /\beta$;
$\beta$ is the electron velocity $\beta=|\vec{p}_e|/E_e$,
and $\gamma_{nn}=\sqrt{m_NE_e^{max}}$.
Furthermore, 
\bea
G_V'^2 &=& (G_F V_{ud})^2
\left[1+\frac{\alpha}{2\pi}e_V^R\right]\, ,
\label{eq;GV}
\eea
where $e_V^R$ is an LEC that 
appears in calculating radiative corrections.
Finally
$\delta_\alpha^{(1)}$ is the radiative correction
of ${\cal{O}}(\alpha)$,
\footnote{
We note that $\delta_\alpha^{(1)}$ is not exactly
the same as the outer radiative correction 
$g(E_0,E)$ given in \cite{s-pr67},
owing to a slightly different renormalization scheme
employed calculating $\delta_\alpha^{(1)}$~\cite{aetal-plb04};
these two quantities are related as  
$\delta_\alpha^{(1)}=g(E_0,E) + 5/4$.
This feature leads to a difference 
of $\frac54\frac{\alpha}{2\pi} \simeq 1.45\times 10^{-3}$
in each of the subsequent expressions.
}
\bea
\delta^{(1)}_\alpha &=& 
3{\rm ln}\left(\frac{m_p}{m_e}\right)
+\frac12
+ \frac{1+\beta^2}{\beta}{\rm ln}\left(\frac{1+\beta}{1-\beta}\right)
-\frac{1}{\beta}{\rm ln}^2\left(\frac{1+\beta}{1-\beta}\right)
+ \frac{4}{\beta}L\left(\frac{2\beta}{1+\beta}\right)
\nnb \\ && 
+ 4\left[\frac{1}{2\beta}{\rm ln}\left(\frac{1+\beta}{1-\beta}\right)-1
\right]\left[
{\rm ln}\left(\frac{2(E_e^{max}-E_e)}{m_e}\right)
+\frac13\left(\frac{E_e^{max}-E_e}{E_e}\right) -\frac32\right]
\nnb \\ &&
+\left(\frac{E_e^{max}-E}{E_e}\right)^2 
\frac{1}{12\beta}{\rm ln}\left(\frac{1+\beta}{1-\beta}\right)\, ,
\label{eq;delta1}
\eea
with $L(x) = \int^x_0 \frac{dt}{t}{\rm ln}(1-t)$.

Eqs.~(\ref{eq;cross-section}) and (\ref{eq;GV})
involve two LECs, 
$e_V^R$ and $l_{1A}$,
which need to be fixed.
The LEC $e_V^R$ can be fixed using the 
experimental value of the neutron lifetime $\tau$ and the 
axial current coupling $g_A$.
We use the expression for $\tau$ 
given in Ref.~\cite{aetal-plb04} 
and employ the experimental values,
$\tau= 885.7(8)$ sec and $g_A=1.2695(29)$
quoted in PDG2004~\cite{PDG2004}. 
The LEC, $l_{1A}$, can in principle be 
fixed by applying dEFT
to the A=3 nuclear systems
and using the tritium $\beta$-decay
rate to constrain $l_{1A}$,
a procedure similar to  
the one adopted 
in the EFT* calculations~\cite{petal-prc03}.
However, a dEFT calculation
for the three-nucleon system
with an external weak current is yet to be done.
We therefore make here partial use
of the results obtained 
in the EFT* calculations~\cite{petal-prc03}. 
According to Ref.~\cite{petal-prc03},
the cross section 
for charged-current weak-interaction processes
in the two-nucleon system
receives about 2 percent corrections
from the (two-body) exchange current;
see, e.g., Eq.~(29) in Ref.~\cite{petal-prc03}.\footnote{
Since each scheme of EFT comes with its specific
definition of the field variables,
the comparison of the strengths of the two-body contributions 
in different schemes involves subtleties.
In the present exploratory study, however,
we simply use the result of EFT* 
to fix the value of $l_{1A}$ in dEFT.
We will give a brief remark on this point
later in the text.
It is worth mentioning here
that the precision of the calculated cross section
turns out to be controlled 
by the errors in the
currently available experimental values of 
$a_0^{nn}$ and $r_0^{nn}$.} 
We may then fix $l_{1A}$ by 
imposing the condition that
the term involving $l_{1A}$ should enhance the 
${\bar{\nu}}_e d\rightarrow e^+ nn$ cross section   
by 2 \% at the initial neutrino energy 
$E_\nu=20$ MeV~\cite{ak-05}. 
This requirement leads to
\bea
\frac{\alpha}{2\pi}e_V^R = (2.01\pm 0.40) \times 10^{-2} \, ,
\ \ \
l_{1A} = -0.33 \pm 0.03 \, .
\eea
Here we have used 
$G_F= 1.16637(1)\times 10^{-5}$ GeV$^{-2}$
determined from muon decay \cite{PDG2004},
and $V_{ud}=0.9738(4)$ deduced from the $0^+\to 0^+$ 
nuclear $\beta$-decays~\cite{ht-04}.
The quoted errors in $e_V^R$
are dominated by the uncertainties in $g_A$,
while the errors in $l_{1A}$ reflect
the $\sim$0.2 \% error in
the calculated $\nu$-$d$ cross sections, 
which in turn are associated with
the errors in the experimental value of 
the tritium $\beta$-decay rate.

We also need to specify 
the $nn$ scattering length 
and effective range.
Their current experimental values are 
(see, e.g., Ref.~\cite{tg-prc87}) 
\bea
a_0^{nn} = -18.5 \pm 0.4\ \ \mbox{\rm [fm]} \, ,
\ \ \ 
r_0^{nn} = 2.80 \pm 0.11 \ \ \mbox{\rm [fm]}\, .
\label{eq;a0r0}
\eea
Thus there are $\sim$2 \% and $\sim$4 \% uncertainties
in $a_0^{nn}$ and $r_0^{nn}$, respectively. 
We will discuss later the consequences
of these uncertainties.
We do not include in Eq.~(\ref{eq;cross-section})
the $1/m_N$ corrections 
or nuclear-dependent (two-body part) 
${\cal{O}}(\alpha)$ corrections, 
the contribution of which are about 0.1 \%.

Having specified the LECs and other parameters
appearing in our formalism,
we are now in a position to calculate
the cross section and electron energy spectrum 
for $nn$-fusion.
We find it convenient to present our
numerical results at a certain specified 
incident neutron energy;
the cross sections at other energies can
be readily obtained by using the $1/v$ law.
So we consider ultra-cold neutrons (UCN).
A typical temperature for UCN is $T_{UCN}\sim 1 $ mK,
and the corresponding average 
velocity is $v_{UCN}\sim 5$ m/sec.
So, for the sake of definiteness,
we may consider a head-on collision 
of two neutrons each moving 
with $v_{UCN}=5$ m/sec;
thus in the CM system of these two neutrons
$v=2v_{UCN}\sim 10$ m/sec.
The numerical results given below
correspond to this kinematics.

\begin{figure}
\begin{center}
\epsfig{file=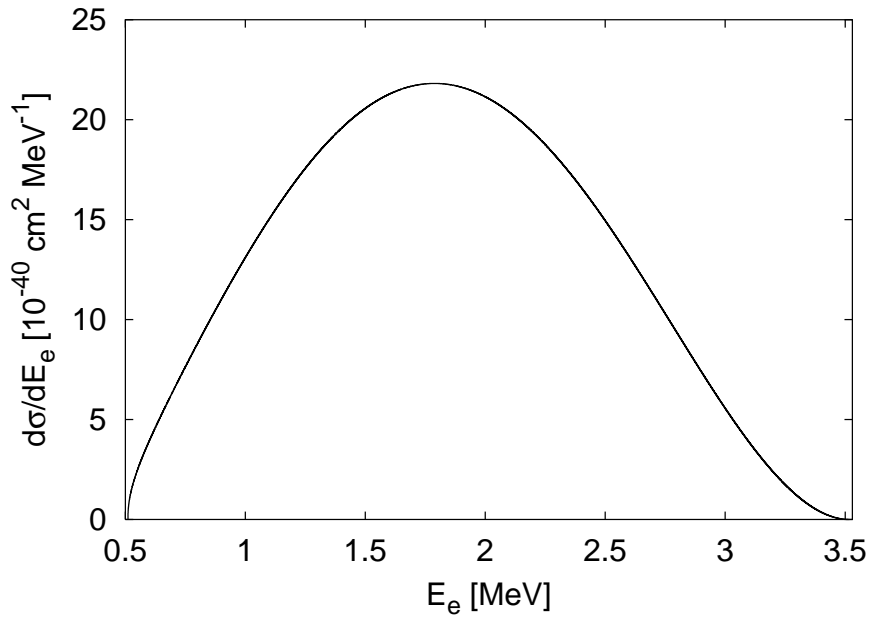,width=12cm}
\caption{Spectrum of the electrons 
from neutron-neutron fusion,
$nn\to de\nu$.}
\label{fig;dsig}
\end{center}
\end{figure}
In Fig.~\ref{fig;dsig},
we plot the calculated electron energy spectrum, 
$d\sigma/dE_e$, as a function of $E_e$.
As mentioned, the electrons with $E_e>\delta_N=1.29$ MeV
can in principle be distinguishable
from the electrons from neutron $\beta$-decay.
Since the amplitude 
in Eq.~(\ref{eq;Amp}) is independent of $E_e$, 
the shape of the electron 
energy spectrum in Fig.~\ref{fig;dsig} is determined 
mainly by the phase factor and the ${\cal{O}}(\alpha)$
corrections coming from the Fermi function and the
radiative correction, $\delta_\alpha^{(1)}$, 
in Eq.~(\ref{eq;delta1}).

We also calculate the total cross section $\sigma$ 
as well as 
$\sigma_{cut}$, the latter being
the differential cross section integrated over 
$E_e>\delta_N$; {\it viz.}
\bea
\sigma &=& 
\int^{E_e^{max}}_{m_e}dE_e\frac{d\sigma}{dE_e}\, ,
\ \ \ 
\sigma_{cut} = 
\int^{E_e^{max}}_{\delta_N}dE_e\frac{d\sigma}{dE_e}\,. 
\eea
The results are 
\bea
\sigma = (38.6 \pm 1.5) \times 10^{-40}\, 
\mbox{\rm[cm$^2$]} ,
\ \ \ 
\sigma_{cut} = (30.2\pm 1.2) \times 10^{-40}\, 
\mbox{\rm[cm$^2$]}\, .
\label{eq;sigmas}
\eea
Since there is no $E_e$-dependence 
in the transition amplitude in Eq.~(\ref{eq;Amp}), 
the relative errors in
$\sigma$ and $\sigma_{cut}$ are the same.
The $\sim$4\% uncertainties in the cross sections 
in Eq.~(\ref{eq;sigmas}) mainly come 
from the errors in
the experimental values of $a_0^{nn}$ and $r_0^{nn}$,
Eq.~(\ref{eq;a0r0}).
We note that the uncertainties in $a_0^{nn}$ 
and $r_0^{nn}$ affect the cross sections
to about the same extent;
the $\sim$ 2.2\% error in $a_0^{nn}$ 
leads to $\sim$ 3.4\% uncertainty, 
and the $\sim$4\% error in $r_0^{nn}$ to
$\sim$1.9\% uncertainty.

\vskip 3mm \noindent
{\bf 4. Discussion and conclusions}

In this paper we studied the $nn$-fusion 
process at low energies
employing the pionless EFT that incorporates
the dibaryon fields (pionless dEFT). 
The electron energy spectrum and 
the integrated cross sections were calculated
up to NLO. 
We included the ${\cal{O}}(\alpha)$ radiative
corrections calculated for the one-body transition
contributions.
Our formalism involves the two LECs, $e_V^R$ and $l_{1A}$.
The former is associated with the inner radiative correction
in $\beta$-decay,
and the latter with 
the short-range two-nucleon electroweak interaction.
We fixed these LECs with sufficient 
accuracy for our present purposes 
in the following manner.
The LEC, $e_V^R$, is fixed using the experimental data 
on neutron $\beta$-decay.  
The LEC, $l_{1A}$, is constrained
with the use of the results obtained 
in the EFT* calculations in the literature,
in which this short-range electroweak effect was
determined from the tritium $\beta$-decay rate.
Once this is done, we can calculate
the $nn$-fusion rate with about 4\% accuracy.
The uncertainties in our theoretical estimates
are dominated by the existing uncertainties 
in the measured values of the $nn$ 
scattering length $a_0^{nn}$ and effective range $r_0^{nn}$.

In view of the enormous difficulty 
of observing $nn$-fusion,
elaborate calculations of its cross section
are not warranted at the present stage.
We have however presented a detailed treatment
of $nn$-fusion, because the same formalism 
can be used 
for the other related processes for which
high-precision calculations are certainly needed.
The remainder of this section is written
in the same spirit. 

To fix the value of $e_V^R$,
we have used here
the current standard values of $G_F$, $V_{ud}$, 
$\tau$, and $g_A$.
We note, however, that 
one of the main purposes
of high-precision measurements of 
neutron $\beta$-decay is 
to deduce the accurate value of $V_{ud}$
avoiding nuclear-model dependence.
It is therefore important to determine $e_V^R$ 
through other experiments.\footnote{
As discussed in Ref.~\cite{fk-appb04}, 
a term (known as the $C$ term) 
in the inner radiative correction~\cite{t-npa92}
has model dependence because,
when the loop diagrams involve the axial current,
one cannot exactly match the calculations 
based on the quark degrees of freedom (in short range) 
with those based on the hadron degrees of freedom 
(in long range).
It seems worthwhile to calculate the $C$ term 
in other models and estimate the model dependence 
of the existing treatments
of the inner radiative corrections.}
Moreover, 
a recent measurement 
of the neutron lifetime $\tau$~\cite{setal-plb05} 
reported a value that differs from the 
existing world average value by 6.5 standard deviations. 
A new precise measurement of 
neutron $\beta$-decay currently under planning 
is essential to clarify 
this discrepancy and to determine 
the value of $V_{ud}$.
(A change in the $nn$-fusion cross section
due to the new experimental value of $\tau$
is smaller than the uncertainties  
due to the limited accuracy of $a_0^{nn}$ and $r_0^{nn}$.) 

We have fixed the value of $l_{1A}$ using the result of
the EFT* and potential model calculations
for the tritium $\beta$-decay.
It is an important future task to determine $l_{1A}$
within the framework of dEFT itself.
For cases that do not involve
external currents,
there exists work in which dEFT is applied 
to the three-nucleon systems~\cite{bhk-npa00}.
We need to extend this type of work
to cases that include external currents.
Meanwhile, with the use of the value of $l_{1A}$ 
deduced in the present work, 
we can carry out dEFT 
calculations of the $pp$-fusion process 
and $\nu d$ reactions 
with no free parameters.
As mentioned earlier, a relation 
between $l_{1A}$ and $L_{1A}$
was discussed in Refs.~\cite{ds-npa04,cijl-05}.
The relation given in Ref.~\cite{ds-npa04} 
leads to $L_{1A}= -0.26\pm 0.11$ fm$^3$,
while that derived in Ref.~\cite{cijl-05}
gives $L_{1A}= 1.18\pm 0.11$ fm$^3$.
The indicated errors
are consistent with those in the value of $L_{1A}$
deduced from the tritium $\beta$-decay, 
$L_{1A}=4.2\pm 0.1$ fm$^3$~\cite{chhr-prc03}, 
but the central values are smaller.
It is possible that this discrepancy stems from 
the different expansion schemes used in these calculations.

We remark that 
the contribution from the $l_{1A}$ term
probably can play a significant role
in other processes,
such as $\pi d \to nn\gamma$ and $\mu d \to nn\nu$,
which are used for deducing $a_0^{nn}$ 
and $r_0^{nn}$.
Recent (pionful) EFT calculations for 
$\pi d\to nn \gamma$~\cite{gp-05} and 
$\gamma d\to nn\pi$~\cite{letal-05} 
indicate that the Kroll-Ruderman (KR) term 
is important for these reactions.
Since the KR term is related to 
the axial-vector couping constant $g_A$ 
through chiral symmetry,
and since $g_A$ is known to be modified 
by the multi-nucleon effects,
it is likely to be important
to include the short-range two-body effects
in analyzing the $\pi d\to nn \gamma$
and $\gamma d\to nn\pi$ processes. 

\vskip 3mm \noindent
{\bf Acknowledgments}

SA thanks 
Y. Li,
A. Gardestig, and 
S.~X. Nakamura for communications.
This work is supported in part by 
the Natural Science and Engineering Research Council of Canada
and by the United States National Science Foundation,
Grant No. 0140214.

\end{document}